\documentclass{article}
\usepackage[T1]{fontenc} % add special characters (e.g., umlaute)
\usepackage[utf8]{inputenc} % set utf-8 as default input encoding
\usepackage{ismir,amsmath,cite,url}
\usepackage{graphicx}
\usepackage{color}

\usepackage{lineno}

\usepackage{subfigure}
\usepackage{algorithm}
\usepackage{algorithmic}
\usepackage{enumitem}
\usepackage{booktabs}
\usepackage{multirow}
\title{MeloForm: Generating Melody with Musical Form based on Expert Systems and Neural Networks}

\multauthor
{Peiling Lu$^1$ \hspace{1cm} Xu Tan$^{\text{*}1}$\thanks{* Corresponding author: Xu Tan, xuta@microsoft.com} \hspace{1cm} Botao Yu$^2$} { \bfseries{Tao Qin$^1$ \hspace{1cm} Sheng Zhao$^3$ \hspace{1cm} Tie-Yan Liu$^1$}\\
 $^1$ Microsoft Research Asia, Beijing, China\\
$^2$ Nanjing University, Nanjing, China\\
$^3$ Microsoft Azure Speech, Beijing, China\\
{\tt\small \{peil,xuta,taoqin,szhao,tyliu\}@microsoft.com, btyu@smail.nju.edu.cn}
}

\def\authorname{F. Author, S. Author, and T. Author}

\begin{document}

\maketitle
%

% MeloForest, MeloForm, MeloForMe, Melorm

%
\begin{abstract}
% The abstract should be placed at the top left column and should contain about 150-200 words.

Human usually composes music by organizing elements according to the musical form to express music ideas. However, for neural network-based music generation, it is difficult to do so due to the lack of labelled data on musical form. In this paper, we develop MeloForm, a system that generates melody with musical form using expert systems and neural networks. Specifically, 1) we design an expert system to generate a melody by developing musical elements from motifs to phrases then to sections with repetitions and variations according to pre-given musical form; 2) considering the generated melody is lack of musical richness, we design a Transformer based refinement model to improve the melody without changing its musical form. MeloForm enjoys the advantages of precise musical form control by expert systems and musical richness learning via neural models. Both subjective and objective experimental evaluations demonstrate that MeloForm generates melodies with precise musical form control with 97.79\% accuracy, and outperforms baseline systems in terms of subjective evaluation score by 0.75, 0.50, 0.86 and 0.89 in structure, thematic, richness and overall quality, without any labelled musical form data. Besides, MeloForm can support various kinds of forms, such as verse and chorus form, rondo form, variational form, sonata form, etc. Music samples generated by MeloForm are available via this link\footnotemark{}\footnotetext{\url{https://ai-muzic.github.io/meloform/}}, and our code is available via this link\footnotemark{}\footnotetext{\url{https://github.com/microsoft/muzic/tree/main/meloform}}.
% \footnote{\url{https://ai-muzic.github.io/meloform/}}
% \footnote{\url{https://github.com/microsoft/muzic/tree/main/meloform}}

\end{abstract}
%

% 统一
% measure - bar
% musical form

\section{Introduction}\label{sec:introduction}

Melody is often composed of hierarchical motifs, phrases and sections with repetitions and variations given musical form \cite{caplin2013analyzing}. This organized structure provided by musical form can help better express music ideas. For example, verse and chorus form is widely used in popular music. The repetition of verse and chorus sections helps emphasize music ideas, while the contrast between verse and chorus can create more emotional intensity. Automatic melody generation with pre-given musical form based on purely data driven technology is difficult due to the lack of labelled data on musical form. Previous work attempt to generate melody with structure information, but still suffers from the following issues: 1) They generate melodies with repetitive patterns implicitly either by learning long-term dependency \cite{wu2019hierarchical,huang2018music,shih2022theme} or representing the repetition structure with same harmony, rhythm patterns, etc \cite{chen2019effect, ju2021telemelody, zhang2021structure}. However, repetitive patterns are far from the exact musical form. 2) They generate melodies with bar-level structure explicitly by learning the relationship between bars\cite{zou2021melons,wu2020popmnet}, but bar-level structure is still not the exact musical form. 3) They generate melodies with repetitive phrases and sections either by detecting phrase labels with rules and algorithms \cite{dai2021controllable} or using human labelling \cite{zhao2021accomontage}. But it is hard for rules or algorithms to detect precise musical form, and it costs much for hiring people to label this musical form data. Furthermore, all of these work aim to generate melodies with some kind of repetitive patterns, but musical form is constructed to express music ideas by developing the hierarchical structural units (i.e., motifs, phrases and sections) with repetitions and variations. Simply repeating some fragments, or even phrases and sections, without considering the relationships among these hierarchical structures, is superficial.

Although it is difficult to collect labeled musical form data through detection algorithm or human labeling, it is much easier for expert systems to generate melodies with musical form. However, experts systems may suffer from monotonous musicality because of handcraft rules. On the other hand, neural networks are capable of creating melodies with rich expressions by learning the data distribution, but it is hard to precisely control the musical form. Considering the complementary characteristics of these two systems, we come up with a method that can leverage the advantages and make up for the shortcomings.

In this paper, we develop MeloForm, a system that generates melody with musical form using expert systems and neural networks. The expert system is designed to generate synthetic melodies with precise musical form. It develops the motifs to phrases then to sections, which are arranged by repetitions and variations according to the pre-given musical form. The encoder-attention-decoder Transformer based neural network is introduced to refine melodies generated by expert systems. To improve musical richness without changing musical form, we propose the refinement strategy in phrase level, the conditioning on rhythm and harmony, and the methods for differentiating sections.

MeloForm enjoys the advantages of expert systems and neural models and avoids their limitations as following: 1) Comparing with expert systems, we can generate melodies with better musical richness. 2) Comparing with the models that implicitly learn the repetitive patterns, we can generate melodies explicitly with precise musical form control. 3) Comparing with the models that depend on bar-level structures, we construct higher-level phrases and sections structures. 4) Comparing with the models using detected phrases labels or human labels, we have the labeled musical form data naturally from expert systems with zero cost and precise accuracy.

The main contributions of this work are as follows:
\begin{itemize}[leftmargin=*]
    \item We develop MeloForm, a system that generates melody with musical form using expert systems and neural networks. This system combines the best of white-box expert system and black-box neural networks for generating melodies with precise musical form and rich melodic expression without any labeled data.
    \item Experimental results demonstrate that MeloForm achieves precise musical form control with 97.79\% accuracy without any labeled data, and outperforms baseline systems by 0.75, 0.50, 0.86 and 0.89 averagely in structure, thematic, richness and overall quality in subjective evaluation.
    \item MeloForm can generate melodies with various kinds of forms, such as verse and chorus form, rondo form, variational form, sonata form, etc.

\end{itemize}

\section{Related work}

Automatic melody generation evolves from grammar or statistical based generation \cite{young2017categorial,quick2013grammar,kikuchi2014automatic,garay2004fugue,wakui2016automatic,takano2014automatic} to deep learning empowered generation \cite{wu2020popmnet, guo2021hierarchical,yu2021conditional, wu2019hierarchical, mishra2019long, yu2020lyrics,yang2017midinet,li2019automatic,ju2021telemelody, sheng2021songmass, huang2020pop, dai2021controllable,katharopoulos2020transformers,hsiao2021compound,zou2021melons,huang2018music,dai2019transformer,wu2020transformer,zhang2022relyme,lv2022re}. In this section, we introduce existing neural networks and expert systems for melody generation with musical structure.

\subsection{Neural Networks for Melody Generation}
Generating structured melody has attracted more attention when modeling long music sequence. Previous work address this problem as following: 1) They implicitly learn the long-term dependency or represent repetitive structure with same musical elements to generate melodies with repetitive patterns. Music Transformer \cite{huang2018music} introduces a relative attention mechanism to capture long-term dependency. Theme Transformer \cite{shih2022theme} proposes a novel gated parallel attention module for generation with theme-based conditioning. Another work \cite{wu2019hierarchical} presents a hierarchical recurrent neural network to model the note-beat-bar structure. Other methods \cite{chen2019effect, ju2021telemelody, zhang2021structure} condition the model with same musical features (e.g., harmony and rhythm patterns) to represent the repetitive structure. 2) They explicitly model the bar-level structure for generating melodies one bar after another. In \cite{wu2020popmnet,jhamtani2019modeling}, the authors leverage the bar related self-similarity matrix to model the relationship between bars for guiding the melody generation. MELONS \cite{zou2021melons} constructs a bar-level structure graph for generating melodies with clear bar-level structures. 3) They collect labeled musical data by detection algorithms for generating repetitions for melodies. Repetitive patterns from melodies are detected by music analysis algorithms in \cite{herremans2016morpheus}, while the boundaries of repetitive phrases are recognized in \cite{dai2021controllable}. All of these works can help realize repetitive patterns for melody in some degree, but they still cannot model precise musical form. 

% In this paper, we propose MeloForm, a system for generating melodies with musical form based on expert systems and neural networks. By leveraging expert systems, we can explicitly control the generated melodies by the precise musical form without any labeled data. By involving neural networks, we can ensure the musical richness without changing the musical form. Combing with these two systems, the generated melodies are well organized by hierarchical structures of motifs, phrases and sections with repetitions and variations, which can better express music ideas.

\subsection{Expert Systems for Melody Generation}

Back to 18th century, a system called music dice game is developed for randomly generating music from precomposed options \cite{hedges1978dice}. Recently, much work still investigate rule-based algorithm compositions for specific purpose. The author in \cite{wiriyachaiporn2018algorithmic} comes up with a rule-based algorithm to generate melody note sequence, which is constructed for comparison with the machine learning based compositions. dMelodies \cite{pati2020dmelodies} combines the designed latent factors to create 2-bar melodies for improving data diversity in disentanglement learning. However, none of them consider the rules for generating melodies with musical form. Computoser \cite{bozhanov2014computoser} proposes a hybrid probability/rule based algorithm for music composition, which based on rules about structure, rhythm, repetition, variations, endings, etc. But it did not provide the methods about how to arrange these elements with musical form. And the method for constructing a phrase by multiple different motifs may brings about divergence from music ideas. In \cite{elowsson2012algorithmic}, repetition of phrases is also considered at a rhythmical level and concerning pitch intervals, but the melody in each phrase is generated one note at a time, which is different from a common composition process by developing a phrase from a motif. Comparing with these systems, the expert system in MeloForm generates melodies by considering the musical form as the hierarchical structure of motifs, phrases and sections with repetitions and variations, which can better express music ideas.

% Afterwards, Lejaren Hiller and Leonard Isaacson \cite{alpern1995techniques} develop a generator/modifier/selector approach for the first-try computer generated composition.
% Only relying on collected data with human labels \cite{wang2020pop909} is not economical efficiently.Much synthetic dataset \cite{xiang2016objectnet3d, xiang2014beyond} is constructed for augmenting image data in computer vision domains. However, there are few works that creates synthetic music dataset for specific tasks. 

\section{Method}

To combine the advantages of precise musical form control by expert systems and musical richness learning by neural networks, we develop MeloForm that is shown in Figure \ref{fig:overview}, which contains two modules: 1) expert systems for generating synthetic melodies with musical forms; 2) Transformer based neural networks for refining the generated melodies from expert systems.

\subsection{Melody Generation with Expert System}

\begin{figure}
 \centerline{
 \includegraphics[width=1\columnwidth]{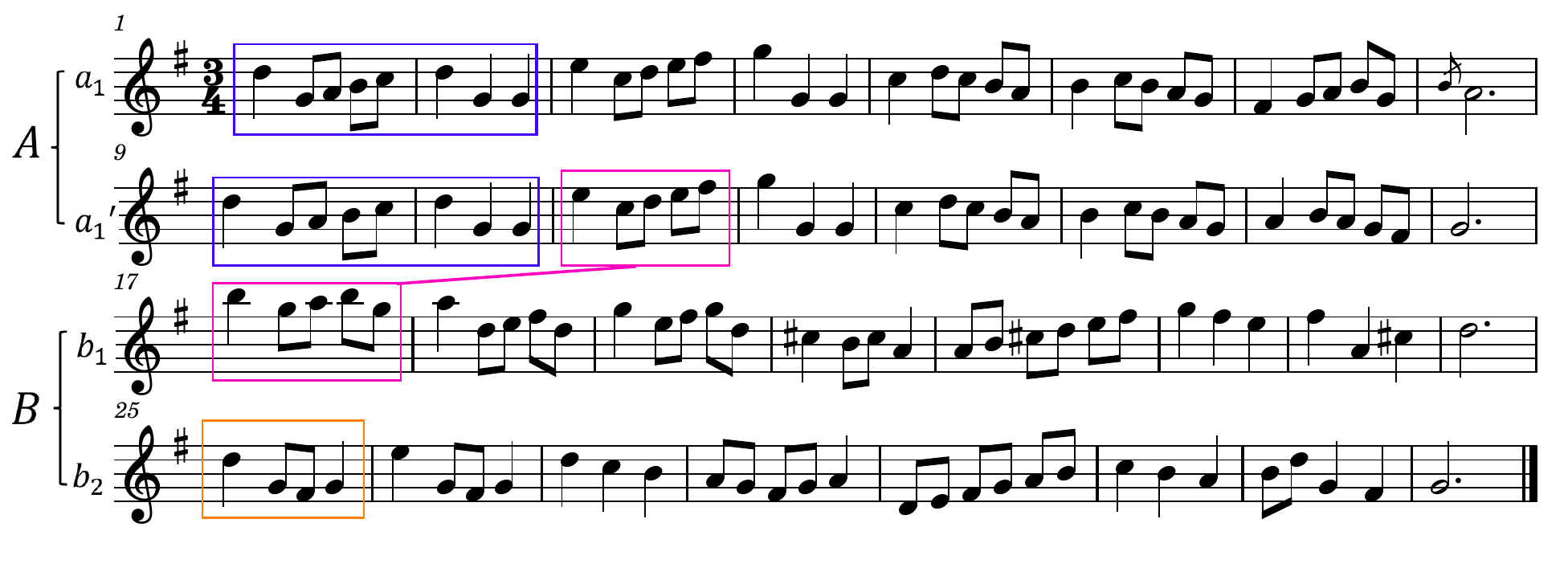}}
 \caption{The score of the melody part from ``\textit{Minuet in G Major}'' by the composer Christian Petzold. This is an example of a melody with musical form as $A(a_{1},a_{1})B(b_{1},b_{2})$. We use capital letters (e.g., $A$, $B$) to label sections, while using lowercase (e.g., $a_{1}$, $b_{1}$, $b_{2}$) for representing phrases. Different phrases in the same section are labeled with different numbers.}
 \label{fig:binary}
\end{figure}

\begin{figure}[t!]
\centering

\subfigure[Pipeline of MeloForm.]{
\label{fig:overview}
\includegraphics[width=1\columnwidth]{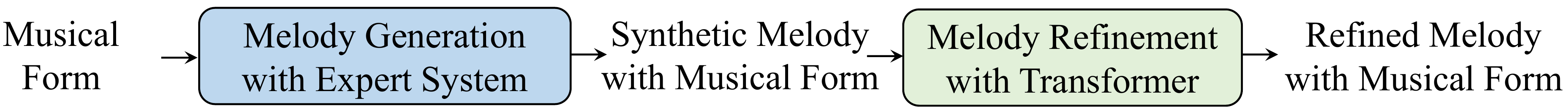}
}
\subfigure[Detailed process of the expert system.]{
\label{fig:expert}
\includegraphics[width=0.99\columnwidth]{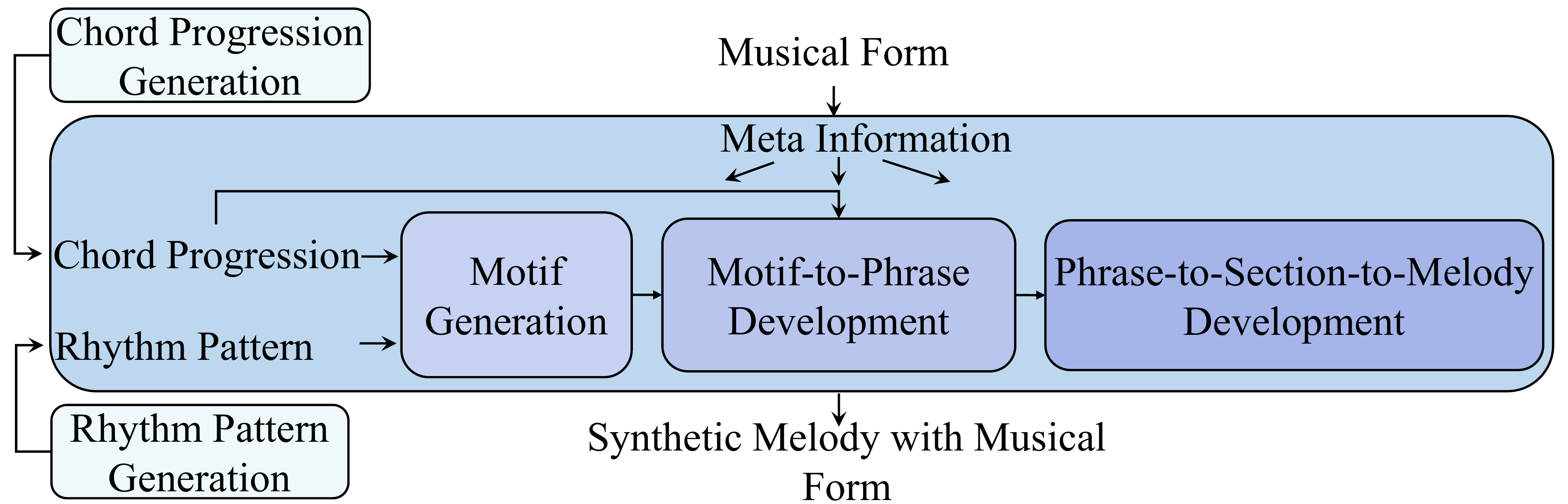}
}
\subfigure[Architecture of melody refinement neural networks. The input contains four phrases from a melody with musical form as $A(a_{1},a_{1})B(b_{1},b_{2})$. In each phrase, $x$ represents the conditioning information (i.e., rhythm, chord and cadence) for each note, while $y$ represents the melody information (i.e., rhythm and pitch) for each note. {[sep]}  indicates the boundaries between phrases, while <s> and </s> indicates the beginning and end of phrases.]{
\label{fig:nn}
\includegraphics[width=0.99\columnwidth]{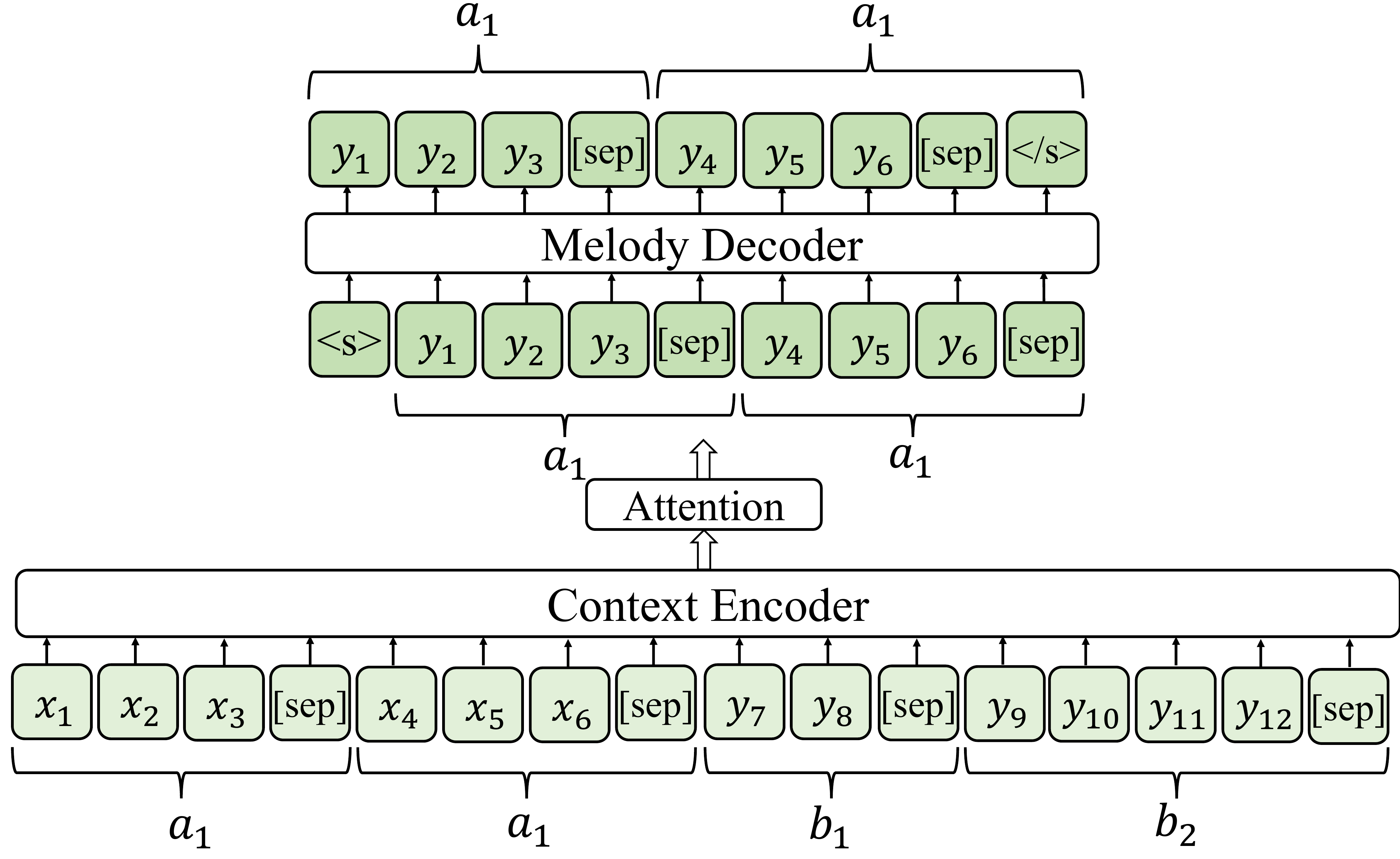}
}
% \vspace{-3mm}
\caption{Architecture of MeloForm.}
\end{figure}

The designed expert system with purely handcraft rules is inspired by music theory \cite{caplin2013analyzing}, and is shown in Figure \ref{fig:expert}. Given the musical form with hierarchical structure of sections and phrases, the expert system firstly generates the motifs based on chord progression and rhythm patterns, then develops the generated motifs to phrases. In the phrase-to-section-to-melody development module, we arrange the phrases in sections and sections in melody with repetitions and variations according to the given musical form to generate the synthetic melody with musical form. For example in Figure \ref{fig:binary}, given the musical form $A(a_{1},a_{1})B(b_{1},b_{2})$, a 2-bar motif in the blue box is developed into 8-bar phrase $a_{1}$. Section $A$ is formed by placing repeated phrase $a_{1}$ sequentially with some variations. The similar process is implemented to form section $B$, in which different phrases $b_{1}$ and $b_{2}$ are developed from different motifs. Then section $A$ and section $B$ are placed sequentially for getting the final composition.

\subsubsection{Motif Generation}

Motif is the smallest structural unit for identifying music theme, so we create an initial melody in a few bars (i.e., usually 1-2 bars) to construct the motif. Before the generation process, we should define the meta information (i.e., scale, pitch range, tempo and meter) for guiding the whole generation procedure. Considering note sequences are composed of pitches and rhythm patterns, we come up with rule-based algorithms for chord progression generation to guide pitch selection and for rhythm pattern generation to create rhythm patterns for the note sequence in motif. Specifically, the initial pitch of notes is selected from tones in corresponding chords. Then we add some embellishing tones$\footnote{\url{http://openmusictheory.com/embellishingTones.html}}$ to decorate the motif melody. Besides, the interval between adjacent note pitches is constrained to be not greater than seven semitones to ensure pitch consistency. The detailed algorithms of chord progression generation and rhythm pattern generation can be referred in Supplementary Materials Section 1 and 2.

% For simplicity, We will illustrate our method in the following with the context of C major scale, pitch ranging from F2 to F4, 60bpm, and 4/4 meter. 
% The motif is generated from scratch as following: 1) Determining the measure length for your motif; 2) Generating the corresponding rhythm patterns and chord progressions; 3) Choosing the note pitch based on the according chord step by step for each position. The pitch should exist in chord tone, and the interval of the pitches from neighboring notes is limited to less than four degrees to ensure consistency for melodic listening. 4) Ornamenting melodies by embellishing tones$\footnote{\url{http://openmusictheory.com/embellishingTones.html}}$ like passing tone (PT), complete neighbor tone (NT), double neighbor figure (DN), etc.

% Melody with too much different motifs may distract listeners from concentrating on the main ideas, so we present another way for deriving similar motif from already generated ones for other phrases. Specifically, instead of generating from scratch, we can either directly copy the generated motif or borrow its rhythm patterns to convey a sense of unity. This mechanism is always used when phrases are belong to the same section for creating a much close connection.

\subsubsection{Motif-to-Phrase Development}

A motif is a unit for identifying the music theme, but it is not a complete music expression. Thus, we need to develop it into a phrase, which is the smallest structural unit for expressing complete music ideas. In order to develop motif into phrase, we introduce three basic categories of development strategies: sequence\footnote{\url{https://en.wikipedia.org/wiki/Sequence_(music)}}, transformation and ending. These basic development strategies can be combined with each other to create compound development strategies. For example, in Figure \ref{fig:binary}, the 2-bar motif in the blue box is developed by sequence (i.e., 3-4 bars), transformation (i.e., 5-6 bars) and ending (i.e., 7-8 bars) to form 8-bar phrase $a_{1}$. The detailed description of each development strategy can be referred in Supplementary Materials Sections 3.

% For example, when developing a motif with one bar to a phrase with four bars, we can firstly restate it to generate sequence for the second bar, and then we restate it again with transformation following to generate the sequence for the third bar, finally we use ending with transformation to create the sequence for the last bar to finish the phrase. 

There may exist multiple different phrases in each section, each of which has its own motif. Too much different motifs in each section may distract listeners from the main ideas. Thus, we present another way to generate a new motif similar to already generated one for a new phrase. Specifically, we can either directly copy the already generated motif or borrow its rhythm patterns to create a new motif, which can help build some relationships between phrases in the same section. For example, in Figure \ref{fig:binary}, the rhythm patterns from the motif in the orange box of phrase $b_{2}$ is borrowed from the motif in the pink box of phrase $b_{1}$ with a few variations.

\subsubsection{Phrase-to-Section-to-Melody Development}

After generating all phrases, we need to arrange them with repetitions and variations to form sections, which are higher-level structural units. And sections are then ordered sequentially to finish the composition. As shown in Figure \ref{fig:binary}, given musical form as $A(a_{1},a_{1})B(b_{1},b_{2})$, repeated phrases $a_{1}$ with variations are placed in order to form section $A$. Phrase $b_{1}$ and $b_{2}$ are placed in order to form section $B$. At last, section $A$ and section $B$ are arranged sequentially to get the final composition.

% The musical form is given as hierarchical structure of sections (e.g. A) and phrases (e.g. a1). For example, we represent the musical form in \ref{fig:binary} as A(a1-a2)-B(b1,b2)-A(a1,a2). Firstly, phrases in sections and multiple sections are placed sequentially in order to conform to the musical form. Secondly, we perform some variations to the exact repeated phrases (e.g. all phrases labeled with a1) by adding embellish tones or fine tune few notes (e.g. splitting, merging or moving notes).
Directly generating new motifs when building different sections is not always musically expressive, so we introduce another method for establishing not only similar but also contrastive motifs. To establish the similarity, we can borrow one of the random fragment from phrases in other sections. To form a contrast, we can adjust the rhythm patterns and pitch selection in desired section to change the tension. For example, in Figure \ref{fig:binary}, the rhythm patterns of the motif in the pink box from phrase $b_{1}$ is borrowed from that of the third bar from phrase $a_{1}$, which is not the exact motif for phrase $a_{1}$. And to form a contrast, the pitch is selected from higher range for more tension.
% Furthermore, since sections is a complete but dependent music structure, we not only need to establish similarity for different sections, but also form a contrast. To establish the similarity, we are inspired by the method for generating similar motif in Section 3.2.2, except that the motif we borrowed is from the random fragment that is not the motif part in the phrase. To form a contrast, we can adjust the rhythm patterns and pitch selection in desired section to change the tension.

% Given the music forms of hierarchical section and phrase structures, the phrases inside a section should be ordered and the sections in the full song need to be arranged sequentially. There are some rules need to be followed for consistency: 1) To ensure the smooth transition between two adjacent phrases, first note pitch from the 2nd phrase needs to be as close as possible with the last note pitch from the 1st phrase by moving the 2nd phrase with octaves. 2) It is optional that you can add some link notes between these two notes for a smoother transition. 3) The transitions between two sections is open-ended, considering they always express different music ideas.

\subsection{Melody Refinement with Transformer}
 
Melodies generated by the expert system are guaranteed to have precise musical form, but are lack of musical richness since they are generated by handcraft rules. Thus, in this section, we introduce an encoder-attention-decoder Transformer that takes the synthetic melodies as input and outputs refined melodies with better musical richness. However, the refinement is challenging since it needs to keep the musical form unchanged while improving the musical richness. Due to the lack of controllability of neural networks, directly refining the whole melody without any constraints is easy to lose musical form information. To address this challenge, we describe several design principles in our Transformer model for refinement: 
\begin{itemize}[leftmargin=*]
\item \textit{Refining melody phrases by phrases.} Since musical phrase is the smallest structural unit for a complete musical expression, it is better to refine the melody phrase by phrase in an iterative process, instead of refining the whole melody in a single time that fails to maintain the musical form from the expert system. However, only refining one phrase at a time is hard to model the repetitive patterns from similar phrases. Therefore, we should refine similar phrases at each iteration step. 
\item \textit{Refining phrases by conditioning on rhythm and harmony.} Directly generating the refined musical phrases from scratch may lose bottom-level music structure determined by rhythm patterns and harmony. Thus, we should provide explicit condition and control with rhythm and harmony to guide the refine process.
\item \textit{Refining phrases by considering their differentiations among different sections.} When composers need to distinguish different sections (e.g., verse and chorus) in melodies, they usually change pitch distribution or rhythms patterns for building up contrasted tension. Accordingly, to maintain such contrast between sections, we need to consider these differentiations in the refinement process.
\end{itemize}

Based on these design principles, our model architecture is shown in Figure \ref{fig:nn}. We introduce the implementations corresponding to each design principle as follows.

%The model architecture is shown in Figure \ref{fig:nn}, which contains a context encoder, a melody decoder and a designed attention module. The context encoder takes in the context information of synthetic melodies from expert systems, while the melody decoder predicts the corresponding refined melodies for specific phrases. Considering the predicted melody of the target phrase is highly correlated to the conditions provided by corresponding source phrase and weakly correlated to the melodies of other phrases, we apply an phrase-level alignment constraint on encoder-decoder attention for helping melody in target phrases only attending to the conditions in corresponding source phrases.

\subsubsection{Iterative Phrase Refinement}
According to the first design principle, we refine the similar phrases in each iterative step. To train the Transformer model with iterative refinement capability, we first mask similar phrases of the melody in the training data and take the masked melody as the input of the encoder, and generate original phrases corresponding to the masked positions with the decoder, like the masked sequence to sequence (MASS) task in~\cite{song2019mass}. After model training, the Transformer model is used to refine the similar phrases in synthetic melodies from expert systems in an iterative way. Specifically, if we want to refine the melody with musical form as $A(a_{1},a_{1})B(b_{1},b_{2})$, we can refine phrases $a_{1}$ for the first iteration and replace the two original phrases with the refined versions, and then refine the phrase $b_{2}$ based on the refined melody. This iteration step is finished until all phrases are refined.

\subsubsection{Condition with Rhythm and Harmony}

The phrases are developed from motif by sequence strategy that results in similar rhythm patterns within some bars. Besides, harmony plays an important role to control pitch distribution in phrase development. Thus, to maintain these music structures in refinement, we condition the model with rhythm patterns, chord progression and cadence to encourage the model to only refine the pitch of the phrase. Specifically, instead of replacing the masked phrases with masked symbols in MASS~\cite{song2019mass}, we replace them with the corresponding rhythm, chord and cadence symbols, shown as $x$ in Figure~\ref{fig:nn}.

\begin{table*}[t]
 \begin{center}

 \begin{tabular}{l l c c c c}
 \toprule
   \empty & \empty  & Structure$\uparrow$ & Thematic$\uparrow$ & Richness$\uparrow$ & Overall$\uparrow$ \\
  \midrule
  \hfil\multirow{2}{*}{Dataset}\hfill & LMD \cite{raffel2016learning}  & 3.18 $(\pm0.64)$ & 3.07 $(\pm0.58)$ & 3.14 $(\pm0.40)$ & 3.00 $(\pm0.57)$ \\

  & POP909 \cite{wang2020pop909}  & 4.06 $(\pm0.49)$ & 3.83 $(\pm0.54)$ &  4.00 $(\pm0.61)$ & 4.11 $(\pm0.46)$ \\
  \midrule
  
  \hfil\multirow{4}{*}{Method}\hfill & Music Transformer \cite{huang2018music}  & 3.00 $(\pm0.76)$ & 3.21 $(\pm0.74)$ & 2.11 $(\pm0.46)$ & 2.32 $(\pm0.54)$ \\

  & MELONS \cite{zou2021melons} & 3.18 $(\pm0.64)$ & 3.07 $(\pm0.58)$ & 2.64 $(\pm0.64)$ & 2.89 $(\pm0.52)$ \\

  & POP909\_lm  & 2.61 $(\pm0.62)$ & 2.93 $(\pm0.60)$ & 2.96 $(\pm0.45)$ & 2.96 $(\pm0.49)$ \\
  
  \cmidrule{2-6}
  & MeloForm  & \textbf{3.68} $(\pm\textbf{0.35})$ & \textbf{3.57} $(\pm\textbf{0.36})$ & \textbf{3.43} $(\pm\textbf{0.43})$ & \textbf{3.61} $(\pm\textbf{0.34})$ \\

  \bottomrule
 \end{tabular}

\end{center}
 \caption{Subjective evaluation results, with mean opinion scores and 95\% confidence interval for each metric.}
 \label{tab:main}
\end{table*}

\subsubsection{Differentiation between Sections}
% 区分section, 除了可以通过乐思，还可以通过控制其他音乐元素改变张力的方式
We differentiate the phrases in different sections into aspects: 1) Controlling rhythm patterns in the expert system. Rhythm patterns can be adjusted directly in expert system by increasing/decreasing note density, prolong/shrink note length, etc. It is common that larger note density or faster tempo can bring about more intensity. 2) Controlling the pitch distribution in the Transformer model. Although expert systems can control the pitch distribution by selecting pitch from different ranges, it is hard for neural networks to control in the same way. Thus, we insert an ``AVGPITCH'' token at the beginning of the condition from each target phrase to represent the average pitch of this phrase, and another ``SPAN'' token to indicate the pitch span (i.e., the difference between the maximum pitch with minimum pitch). Higher average pitch and more pitch span can build up much more tension.

%Considering composers sometimes need to distinguish different sections (e.g., verse and chorus) in melodies for building up contrasted tension, we introduce two ways for this purpose based on MeloForm: 1) controlling rhythm patterns in expert systems. Rhythm patterns can be adjusted directly in expert system by increasing/decreasing note density, prolong/shrink note length, etc. It is common that larger note density or faster tempo can bring about more intensity. 2) controlling the pitch distribution by neural networks. Although expert systems can control the pitch distribution by selecting pitch from different ranges, it is hard for neural networks to control in the same way. Thus, We insert an ``AVGPITCH'' token at the beginning of the condition from each target phrase to represent the average pitch of this phrase, and another ``SPAN'' token to indicate the pitch span (i.e., the difference between the maximum pitch with minimum pitch). Higher average pitch and more pitch span can build up much more tension.

\section{Experimental Results}
In this section, we first describe the dataset and system configurations. Next, we show the main results compared with baseline systems. Then we implement some method analysis to further validate the effectiveness of each design. Finally we implement some extensions to demonstrate the scalability of this system. 
% Music samples generated by MeloForm are available via this link\footnotemark{\label{}}. Our code and Supplementary materials are available via this link\footnotemark[\value{footnote}].

\subsection{Experiment Setup}

\noindent\textbf{Dataset.} We utilize the LMD-matched MIDI dataset \cite{raffel2016learning} in the training stage of the Transformer based neural networks for melody refinement. We select melodies in 4/4 time signature, and normalize them to the tonality of ``C major'' or ``A minor''. At last, we obtain 30,218 MIDI samples. This dataset contains 471,058 phrases, in which there exists 100,948 distinct set of similar phrases based on our phrase boundary detection and similarity calculation, of which the detailed algorithm can be found in Supplementary Materials Section 4.

\noindent\textbf{System Configurations.} The encoder-attention-decoder Transformer has 4 encoder and decoder layers, both of which contains 4 attention heads. The hidden size is set as 256. We use Adam optimizer \cite{kingma2014adam} with the learning rate is 0.0005. The dropout rate is set as 0.2 during training. There contains maximum of 4,096 tokens in each batch. The dataset is randomly splitted into training/valid/test set with the ratio of 0.8, 0.1, and 0.1, respectively. We use nucleus sampling \cite{holtzman2019curious} with the p set as 0.9.

\noindent\textbf{Subjective Evaluation Metrics.} For subjective listening, we invite 10 participants to evaluate 30 samples. There are five randomly selected samples from MeloForm and baselines described in Section 4.2. For each listener, they are randomly ordered to avoid perceptual bias and habituation effect. The rating is based on five-point scale. We define the following four metrics: 1) Structure: Does this melody have complete structure with repetitions and variations? 2) Thematic: Can you figure out the musical theme? 3) Richness: Is the melody similarly melodious as human compositions? 4) Overall: Is the overall quality of the melody the same as human compositions?

% \noindent\textbf{Objective Evaluation Metrics} For evaluating the controllability, we leverage the metrics of TA, CA and AA in TeleMelody \cite{ju2021telemelody} to verify the control accuracy of tonality, chord, and cadence condition. And we introduce another BARA, POSA, AVGA, and SPA for calculating the controlling accuracy of rhythm, average pitch, pitch span in condition. We also leave some space for AVGA and SPA by floating up/down by three semitones, since it is difficult for model to accurately match the specific values. To verify the control of musical form, we propose REPA, which is obtained by calculating the similarity score between refined melodies from same phrases.

% To verify if the masking strategy for similar phrases is effective for generating similar melodies for same phrases, we propose REPA, which is obtained by calculating the similarity score between generated phrases in each iteration, and regarding it accurate when similarity score is greater than 0.9, which is consistent with training.

% \begin{table*}[t]
 
%   \centering
   
%   \begin{tabular}{l c c c c c c c}
%   \toprule
%   \empty Metric
%  \empty & AVGA & SPA\\
%   \midrule
%   Ground Truth & 100.00 & 61.74  &  100.00 & 100.00 & 100.00 & 100.00 & 100.00 \\

%   MeloForm & 72.15  & 63.86 & 99.80 & 99.75 & 99.75 & 92.35 & 99.71 \\
%   \bottomrule
%   \end{tabular}
%  \caption{Accuracy Results ($\%$) on model controllability. The closer that TA, CA, BARA, POSA, AA, AVGA, SPA are to Ground Truth, the better the controllability is.}
%   \label{tab:control}
% \end{table*}

\subsection{Main Results}

We compare MeloForm with the following baselines: 1) Music Transformer \cite{huang2018music}, an end-to-end neural network solution which introduces the relative attention to learn long-term repetitive patterns; 2) MELONS \cite{zou2021melons}, a Linear Transformer and structure graph based framework for generating melody with bar-level structure; 3) POP909\_lm, a language model implemented by ourself for leveraging the phrase labels in POP909 \cite{wang2020pop909} to generate melodies given musical form, whose details can be found in Supplementary section 6. These baselines share the same system configurations with MeloForm for fair comparison.
% Specifically, we trained the language model using Transformer as the backbone. The input sequence is encoded the same with MeloForm except that the phrase label is added for each note. When predicting melodies given musical form, we enforce the model to place the given label at the beginning of the phrase, which conditions the model to predict the note sequence for the given phrase. These baselines share the same system configurations with MeloForm for fair comparison.

Table \ref{tab:main} shows the subjective evaluation results in the mean opinion scores (MOS) and 95\% confidence interval for these four metrics over all experimental groups. Comparing with baseline systems, MeloForm outperforms all of them in thematicness, structureness, richness and overall quality. We calculate the controlling accuracy of phrase labels for POP909\_lm to verify if same phrase labels results in similar melodies. The high accuracy of 100\% demonstrates it is capable of generating corresponding repetitive melodies for same phrase labels. It is worth noticing that although POP909\_lm can realize the repetitions of phrases, it still fails to provide good listening experience comparing with MeloForm. This demonstrate simply repeating phrases is still deficient. Besides, the less 95\% confidence interval of MeloForm provides more confidence for its better performance. We also compare MeloForm with human composed data from the LMD-matched MIDI dataset and POP909. MeloForm shows better performance compared with LMD-matched MIDI dataset, since the samples from LMD-matched MIDI dataset are collected from various kinds of publicly-available sources on the internet, which is hard to guarantee the production quality. In contrast, POP909 is constructed by hiring professional musicians for creating the samples and reviewing the whole generation process, which ensures the quality to be the same as realistic compositions. The comparing results with POP909 reveal that there still exist gaps between melodies generated by MeloForm with the human composed ones.

\subsection{Method Analysis}

\noindent\textbf{Controllability Study.} We calculate the controlling accuracy of musical form by calculating the similarity score between generated melodies from same phrases. The accuracy of 97.79\% indicates MeloForm can generate melodies with precise musical form control. We also calculate the accuracy to verify if the pitch distribution is well controlled by average pitch and pitch span. The accuracy of 92.5\% in average pitch controlling and 99.71\% in pitch span controlling demonstrate its controllability to differentiate sections in pitch dimension.

\begin{figure}
 \centerline{
 \includegraphics[width=1\columnwidth]{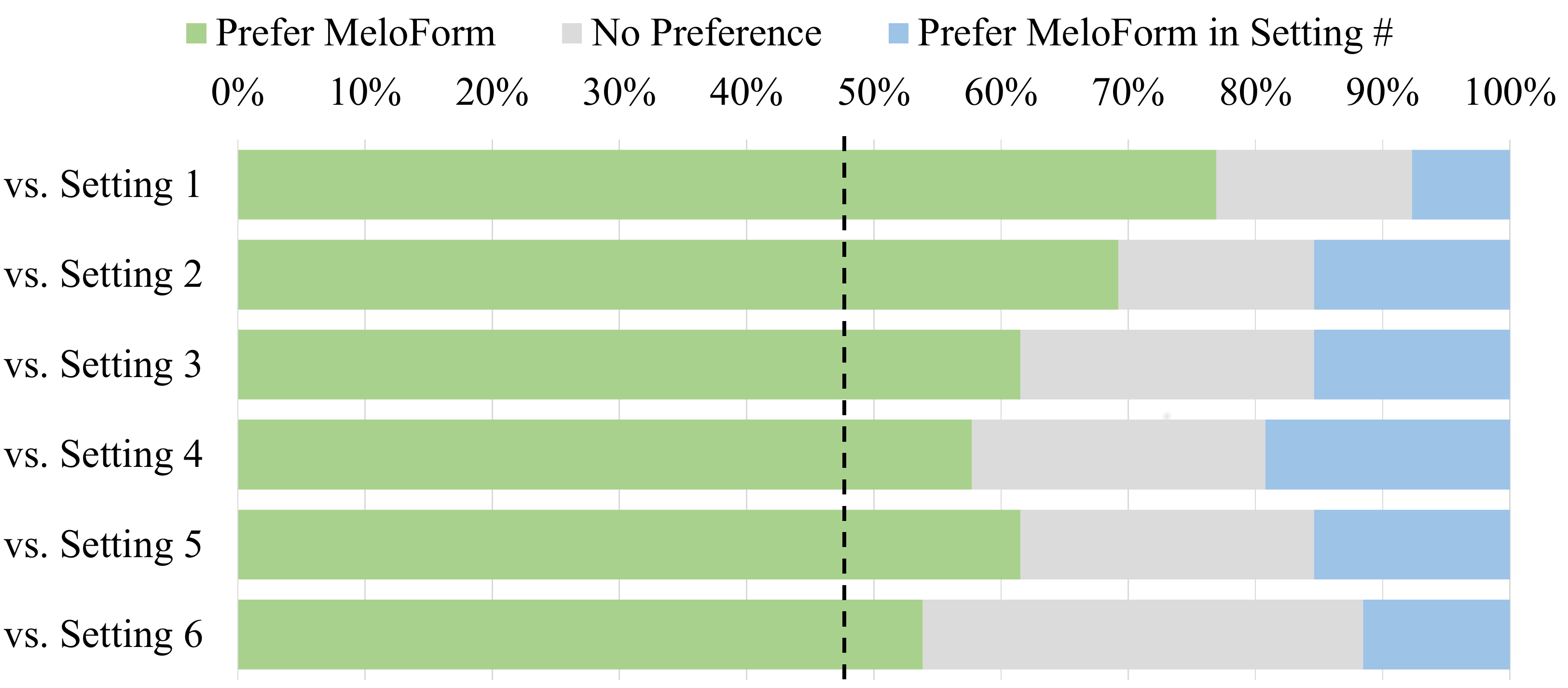}}
 \caption{Preference distribution for ablation study, which compares the melodies generated by MeloForm with that from modified system (i.e., MeloForm in Setting \#).}
 \label{fig:preference}
\end{figure}

\noindent\textbf{Ablation Study.} We conduct preference tests to verify the contribution of the components in MeloForm with the following settings: 1) Setting 1: w/o development strategies from expert systems. Notes are randomly arranged in expert systems without any development strategies. 2) Setting 2: w/o expert systems. Melodies are first generated for each phrase by neural networks, and then got copied directly with the given musical form for repetition in the same phrase. 3) Setting 3: w/o neural networks. We remove melody refinement neural networks. 4) Setting 4: w/o fine-grained rhythm pattern condition. The condition of rhythm patterns changed from fine-grained level to coarse-grained level. 5) Setting 5: w/o refinement strategy. We directly copy the generate melodies from each phrase to the following same phrases. 6) Setting 6: w/o section differentiation. The ``AVGPITCH'' and ``SPAN'' tokens are removed from the beginning of each phrase. The participants are required to compare the samples from MeloForm and these settings and give their preference score.

The preference distribution for the above settings is shown in Figure \ref{fig:preference}. We derived some observations in the following: 1) More preferences over setting 1 and 2 demonstrate the effectiveness of the whole expert system and the development strategies for building up phrases. 2) More preferences over setting 3, 4, 5, and 6 validate the efficacy of the whole neural refinement model, the rhythm pattern condition, the iterative refinement strategy, and the section differentiation method.

\subsection{Extensions to More Musical Forms}

In this section, we illustrate the principles to generate melodies with four different musical forms (i.e., Verse and Chorus, Rondo, Variational and Sonata form) based on MeloForm. The examples from these extensions are shown in the demo page.

\noindent\textbf{Verse and Chorus Form} is widely used in popular music. It contains two contrasting sections: verse section (i.e., $A$) and chorus section (i.e., $B$), which can be arranged in various kinds of ways, such as $AABAAB$, $ABA$, or $AABB$. Users can follow the method in Section 3.1 to build the verse and chorus sections, and leverage the section differentiation in Section 3.2.3 to change the tension. For example, Chorus tends to have more intensive emotional expression. To achieve this, we can control the average pitch to higher level or increase the pitch span for this section.

\noindent\textbf{Rondo Form} is a musical form where the refrained section alternates with contrasting sections, such as $ABACADA$. Melody of Rondo Form can be generated by leveraging the method in Section 3.1 to construct the refrained section and contrasting sections, and arrange them sequentially based on the given Rondo Form.

\noindent\textbf{Variational Form} is a musical form where the main section is followed by its variations. We represent Variational Form as $AA'A''$ to describe the main section and its variations. To address the challenge of deriving variations from the main section, we can treat the main section as a motif, and use the development strategies in 3.1.2 to develop this longer motif into another higher-level section. Another way is to generate melodies with same sections like $AAA$ by expert systems, and predict the melodies from each section in different iteration steps. There are many other ways for creating the Variational Form, so users are encouraged to provide their own creative design.

\noindent\textbf{Sonata Form} is generally consisted of three sections: an exposition, a development, and a recapitulation, which can be represented as $ABA'$ in our system. For the exposition, we generate two phrases with contrasting motifs, where the second phrase is a transposition of the first one by controlling the tonality token when refining the second phrase. The development section is a variation of the exposition, which can be generated by leveraging the methods for constructing variational form. The recapitulation is a repetition of the exposition, in which the second phrase should go back to the tonic key by controlling the tonality token in condition from neural networks.

% \subsubsection{Lyric-to-Melody Extensions}

% Given lyrics with structures of paragraphs and sentences, in which paragraphs are labeled with Verse and Chorus, we develop a method for generating melodies with musical form to the given lyrics. We can follow the same procedure in Section 4.4.1 to build the Verse and Chorus form. Another challenge is to generate melodies in phrases to match the sentences from lyrics automatically. We can equally divided the number of words in sentences into the phrases with desired length, and rhythm pattern is generated based on the number of words in bars of motif. xxx

\section{Conclusions}

In this paper, we propose MeloForm, a system to generate melody with musical form based on expert systems and neural networks. It combines the advantages from expert systems to precisely control musical form and neural networks to refine melody for better musical richness without changing musical form. Experimental results demonstrate MeloForm achieves 97.79\% accuracy in musical form control, and outperforms baseline systems in structure, thematic,richness and overall quality in terms of subjective evaluation. We will release the dataset of the synthetic melodies generated by our expert system and the refined version from our Transformer model to facilitate the future research on music form modeling. Furthermore, we will explore the generation of intro, outro and bridge, and investigate the arrangement generation with musical form to accomplish a complete music composition.

% For bibtex users:
\bibliography{ISMIRtemplate}

% For non bibtex users:
%\begin{thebibliography}{citations}
% \bibitem{Author:17}
% E.~Author and B.~Authour, ``The title of the conference paper,'' in {\em Proc.
% of the Int. Society for Music Information Retrieval Conf.}, (Suzhou, China),
% pp.~111--117, 2017.
%
% \bibitem{Someone:10}
% A.~Someone, B.~Someone, and C.~Someone, ``The title of the journal paper,''
%  {\em Journal of New Music Research}, vol.~A, pp.~111--222, September 2010.
%
% \bibitem{Person:20}
% O.~Person, {\em Title of the Book}.
% \newblock Montr\'{e}al, Canada: McGill-Queen's University Press, 2021.
%
% \bibitem{Person:09}
% F.~Person and S.~Person, ``Title of a chapter this book,'' in {\em A Book
% Containing Delightful Chapters} (A.~G. Editor, ed.), pp.~58--102, Tokyo,
% Japan: The Publisher, 2009.
%
%
%\end{thebibliography}

\end{document}

\maketitle

\section{Chord Progression Generation}
\label{sec:chord}

\begin{figure}
 \centerline{
 \includegraphics[width=1.0\columnwidth]{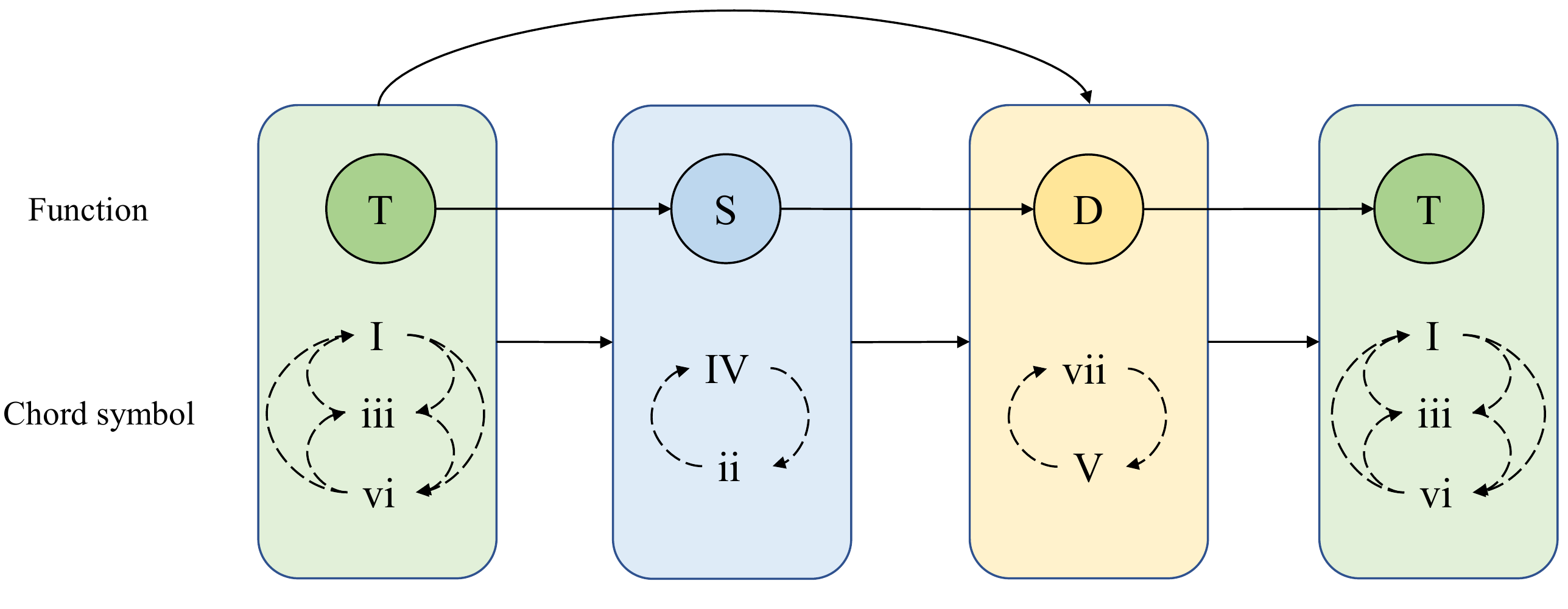}}
 \caption{Transformer based Melody Polisher}
 \label{fig:harmonic}
\end{figure}

The chord progression generation is to generate chord progression for motif and phrase for helping build note sequences in melodies. The chord progression is built upon harmonic functions$\footnote{\url{https://musictheory.pugetsound.edu/mt21c/HarmonicFunction.html}}$, which can be classified as tonic (T), subdominant (S) and dominant (D) in common-practice music. Figure \ref{fig:harmonic} illustrates the harmonic progression flowchart in major scale. The high-level harmonic progression flows as T-S-D-T. In each functional group, the chords can progress to each other as shown in dotted line. The progression for the motif and phrase can be constructed with this pattern. Particularly, at the end of each phrase and sections, cadence$\footnote{\url{https://en.wikipedia.org/wiki/Cadence}}$ is considered when building chord progressions. There are two types of cadences we used: authentic cadence (i.e. chords ending with dominant to tonic chord) and half cadence (i.e. chords ending with tonic chord). Phrases at the end of sections can end with authentic cadence, while phrases inside sections can end with both of them. This depends on how you develop your music ideas. Furthermore, to narrow down the selection range, we collected some best-seller progressions to construct a n-gram distribution for helping direct the chord progression.

\section{Rhythm Pattern Generation}
\label{sec:rhythm}
The rhythm pattern generation creates the rhythm patterns for the motif. For example, We randomly generate the rhythm patterns for one-measure motif in 4/4 meter as follows: 1) We divide the 4-beat measure into uniform eight positions (i.e. one quarter note is partitioned into two eighth notes) as one chunk. 2) We randomly select a subset of this eight positions to insert $n$ rest notes, which divides the original chunk into $n + 1$ chunks. 3) In each chunk, we randomly select the subset of the positions to insert boundaries to determine which set of notes need to be merged. The notes between two consecutive boundaries are bound to be joined into a longer note. There are some limitations to make the rhythm applicable for singing: 1) The accumulated duration of rest notes in one measure is not greater than two beats. 2) The duration of each rest note is not greater than one beat. 3)  The duration of each note is not greater than four beats. 4) The number of notes in each measure is not less than three.

\section{Development strategy}

In order to develop the motif into the phrase, we define three categories of development strategies: sequence, transformation and ending. We will illustrate the specific methods used in each strategy.

\noindent\textbf{Sequence} we leverage the sequence$\footnote{\url{https://en.wikipedia.org/wiki/Sequence_(music)#Melodic_sequences}}$ to restate the motif in development bars, such as real sequence, tonal sequence, rhythmic sequence and modified sequence. The specific one is chosen by referring to the chord in development bars. For example, if the chord progression of development bars is the same as motif, we can randomly choose the methods above without any limitations. However, if the chord progression changes, pitch selection should also be changed for harmony, rhythm sequence is a better choice for this situation.

\begin{figure}[t!]
\centering

\subfigure[Acceleration]{
\label{fig:acceleration}
\includegraphics[width=1\columnwidth]{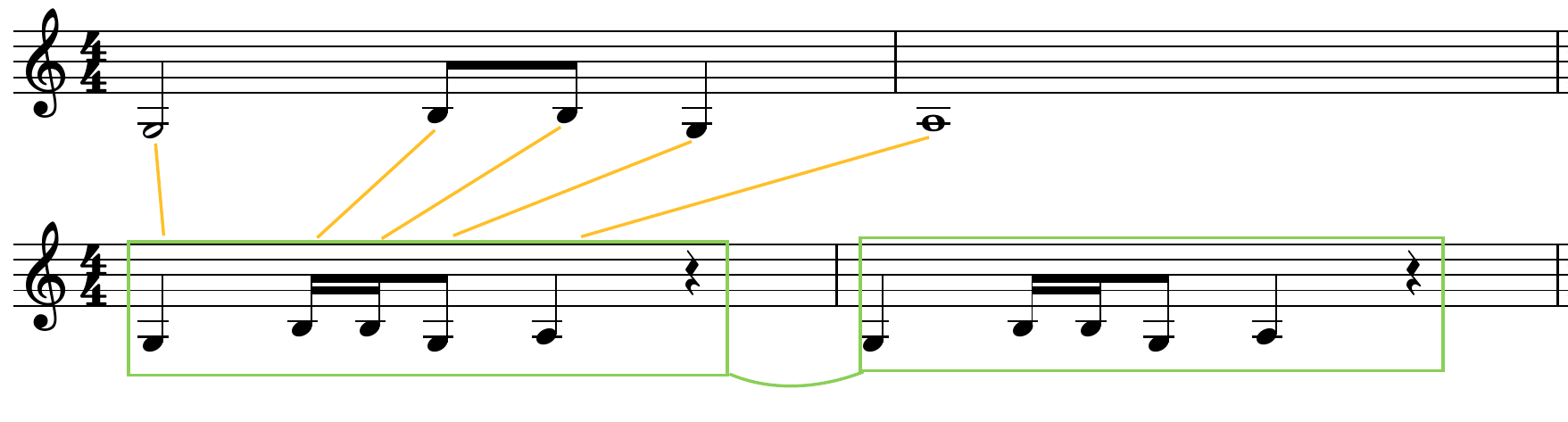}
}
\subfigure[Decoration]{
\label{fig:ornament}
\includegraphics[width=1\columnwidth]{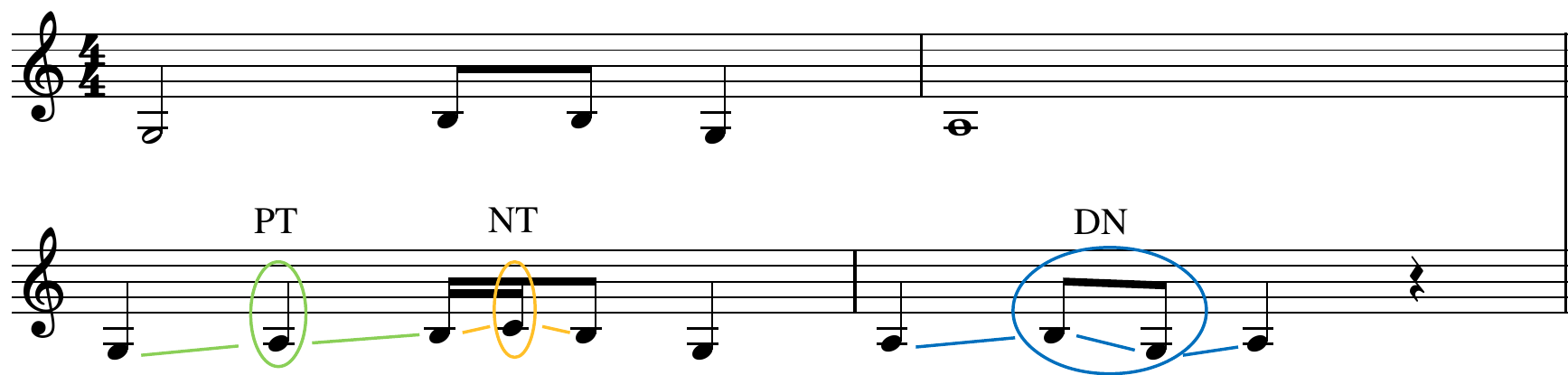}
}

\subfigure[Fragmentation]{
\label{fig:frag}
\includegraphics[width=1\columnwidth]{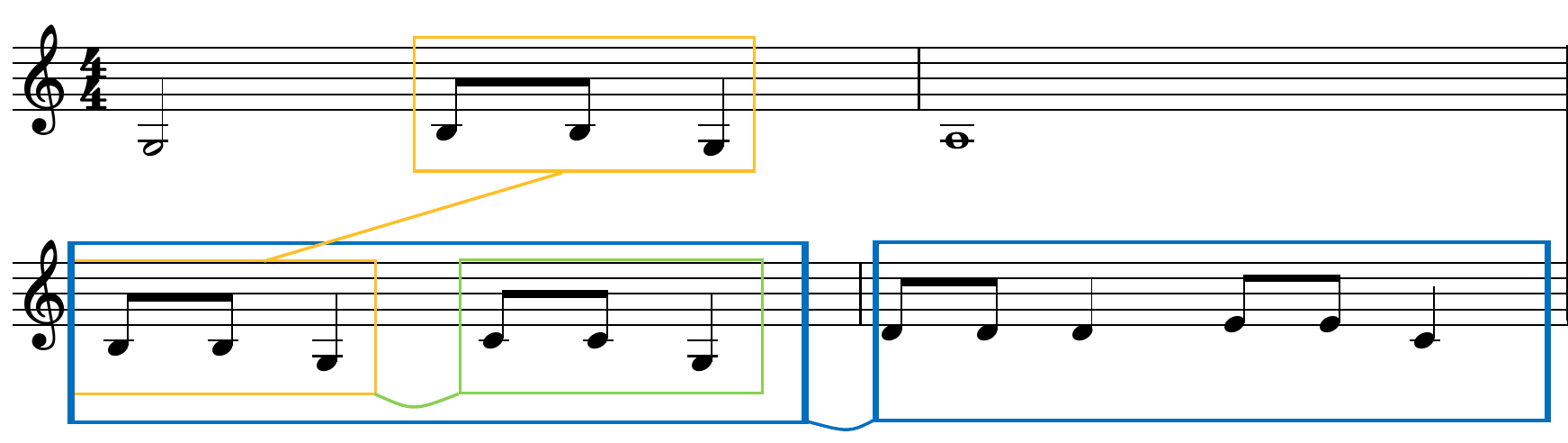}
}
\subfigure[Fine-tuning]{
\label{fig:tune}
\includegraphics[width=1\columnwidth]{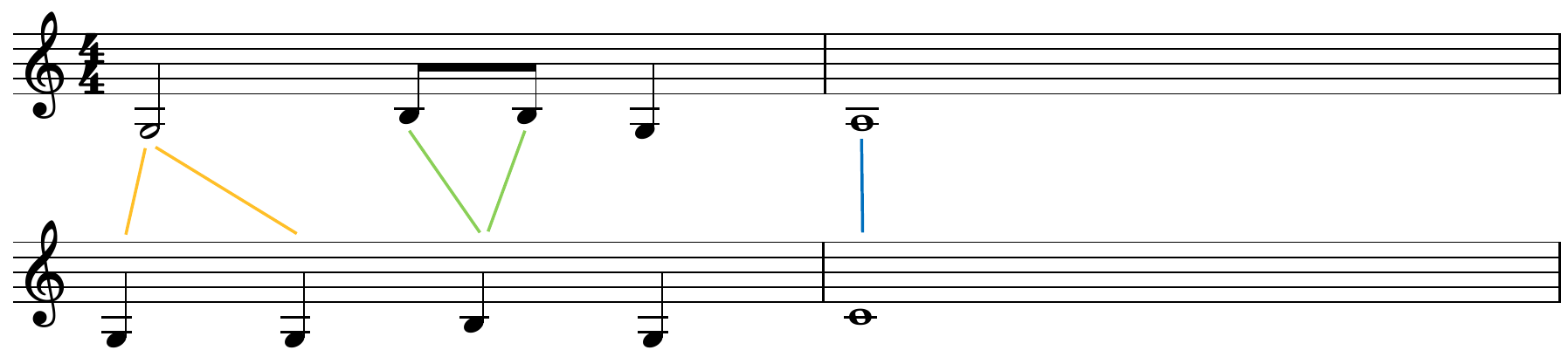}
}
% \vspace{-3mm}
\caption{Transformation strategies. Acceleration: lines in yellow describe the length shortening for each note, while yellow lines illustrate sequence to the next bar. Decoration: yellow notes represent Passing Tone, blue notes represent Complete Neighbor Tone, and blue notes represent Double Neighbor Figure. Fragmentation: motif in yellow box is extracted to be the fragment to be developed. This fragment is firstly sequenced to fill up the remaining bar, and the note sequence from this bar will be sequenced to the next bar for matching the number of bars in original motif. Fine-tuning: yellow lines represent the way to split the notes; green lines represent the method for merging notes; blue lines represent moving pitch.}
\end{figure}

\noindent\textbf{Transformation} For transformation, we propose some methods of acceleration, decoration, fragmentation, and fine-tuning. As shown in Figrue \ref{fig:acceleration}, Acceleration speeds up the motif by reducing note duration and inter-onset interval (IOI)$\footnote{\url{https://en.wikipedia.org/wiki/Time_point}}$. This may results in bar count mismatch since notes are move ahead, sequence can help fill the remaining measures. Decoration in Figure \ref{fig:ornament} is the same with that in motif generation. Fragmentation in Figrue \ref{fig:frag} is to randomly select a fragment of the motif, and sequence it to the desired length for emphasizing the motif once again. Fine-tuning is to change some notes for motif variation by moving the pitch, merging notes, splitting notes, decorating notes, etc.

\noindent\textbf{Ending} Ending is to give a sense of the end, which can be achieved by generating a new melody using the same method of motif generation, and force the last pitch lower than the first pitch to form a downward melody shape. Another way is to prolong the note length in the last bar for a feeling of rest.

\section{Phrase Boundary and similarity Detection Algorithm}
The phrase boundary has been detected by calculating the onset intervals between adjacent notes. If this onset interval is greater than 1.5 beats, and there exists a rest between notes, we recognize there exists a boundary between these two segments.
The phrase similarity is calculated by string edit distance. Saying that there are token sequences of two phrases, we use the string edit distance to calculate the similarity degrees, then regard two phrases are similar if the degree is greater than 0.9.

\begin{figure}
 \centerline{
 \includegraphics[width=1.0\columnwidth]{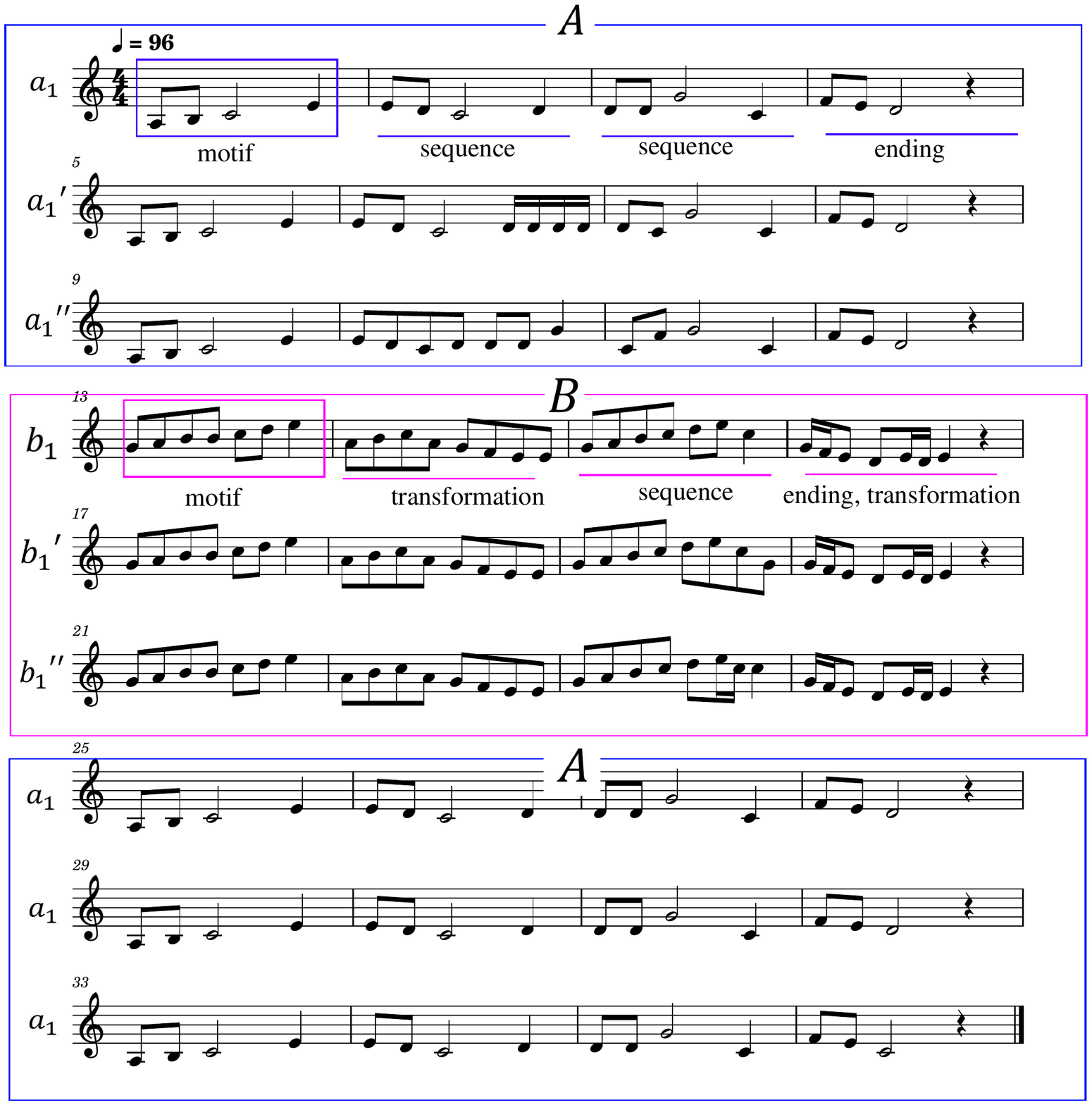}}
 \caption{Case study example.}
 \label{fig:case}
\end{figure}

\section{Case Study}

Figure \ref{fig:case} shows an example of a generated melody from MeloForm. The musical form for this melody is represented as $A(a_{1},a_{1}^{'},a_{1}^{''})B(b_{1}, b_{1}^{'}, b_{1}^{''})A(a_{1},a_{1},a_{1})$. In phrase $a_{1}$, 1-bar motif is developed into 4-bar phrase by sequence, sequence and ending. Combining with phrases $a_{1}^{'}$ and $a_{1}^{''}$, the variations from phrase $a_{1}$, we form the section $A$. For phrase $b_{1}$, motif is developed by transformation, sequence, and a compound of ending and transformation. Phrase $b_{1}$ and its variations $b_{1}^{'}$ and $b_{1}^{''}$ form section $B$. Section A recurs again with three phrases $a_{1}$. More examples are uploaded to demo page$\footnote{\url{https://meloform.github.io/}}$.

\section{POP909\_lm}
We trained the language model using Transformer as the backbone. The input sequence is encoded the same with MeloForm except that the phrase label is added for each note. When predicting melodies given musical form, we enforce the model to place the given label at the beginning of the phrase, which conditions the model to predict the note sequence for the given phrase.